 \documentclass[preprint2]{aastex}

\slugcomment{Accepted for ApJ}

\shorttitle{Magnetic braking of stellar cores  in red giants and supergiants }
\shortauthors{Andr\'e Maeder and Georges Meynet}

\begin{document}


\title{Magnetic braking of stellar cores  in red giants and supergiants }


\author{Andr\'e Maeder\altaffilmark{1} and Georges Meynet\altaffilmark{1}}
\affil{Geneva Observatory, Geneva University, CH--1290 Sauverny, Switzerland}

\email{andre.maeder@unige, georges.meynet@unige.ch}



\begin{abstract}
Magnetic configurations, stable on the long term, appear to exist in various evolutionary phases, from Main-Sequence stars to white dwarfs and neutron stars. The large scale 
ordered nature of these fields, often approximately dipolar, and their scaling according to the 
flux conservation scenario favor the model of a fossil field \citep{Duez2010}.
We make some first estimates  of the
magnetic coupling between the stellar cores 
and the outer layers in red giants and supergiants. Analytical expressions of the
 truncation radius of the field coupling are established for a convective envelope 
and for a rotating radiative zone with horizontal turbulence.  The timescales of
the internal exchanges of angular momentum are considered. 

Numerical estimates are made on the basis of recent model grids. %The magnetic coupling between the core and outer layers is found more efficient in later evolutionary phases. 
 The direct magnetic coupling of the core to the extended convective envelope 
of red giants and supergiants appears unlikely. However, we find that the intermediate radiative zone is fully 
coupled to the core during the He-burning and later phases. This coupling   is able to produce a  strong spin down 
of the core of red giants and supergiants, also leading to relatively  slowly rotating stellar remnants, like 
white dwarfs and pulsars. Some angular momentum is also transferred to the outer convective envelope of red giants and
supergiants  during the He-burning phase and later.
\end{abstract}





\section{Introduction}

The degenerate objects in the final stages of stellar evolution have strong magnetic fields.
Conservation of the  magnetic flux of Main Sequence stars with fields  in the range of 10 G to 1 kG
would lead to neutron stars with magnetic fields in the range of $10^{11}$ to $10^{13}$ G and to white dwarfs
with fields in the range of $10^5$ to $10^7$ G \citep{Chanmugam92}. 
This is generally consistent with the  field of classical pulsars  of the order of $10^{12}$ G.  Recent results
\citep{Landstreet2012} indicate that about 10\% of the white dwarfs (WD) are hosting magnetic fields
in the range   of $10^5$ to  $10^8$ G. At present, it is not clear whether all WD have a magnetic field at a lower level.

The action of magnetic fields is not necessarily limited to the early and final evolutionary stages.
 The fields may be acting in various ways during star evolution, for example  by disk locking during the accretion phase in star formation \citep{Hartmann98}, by magnetic braking due to the stellar winds  of solar type stars \citep{Kawaler88} as well as of massive stars \citep{MEM2011}, by influencing the distribution of internal rotation and mixing of the 
chemical elements \citep{magn03,Heger2005}, by their interaction with meridional circulation, turbulence and shears  \citep{MathisZ2005}, by the viscous and magnetic coupling of the core and envelope \citep{Spada2011}, by producing the pulsar emission and spin--down, etc. 

The long-term stability of non-force-free magnetic topologies has been shown by 
 \citet{Duez2010}.  According to these authors, this might help   to understand some
 stable magnetic configurations existing in various evolutionary phases, from Main-Sequence stars to white dwarfs and neutron stars. Duez et al. point out that the large scale 
ordered nature of these fields, often approximately dipolar, and their scaling according to the 
flux conservation scenario favor a fossil field. Although pure poloidal or toroidal fields are unstable,
 an axisymmetric mixed configuration may provide a stable  equilibrium topology \citep{Tayler73,Markey73,Tayler80}.

The object of the present study is to study the effects of
a dipolar  magnetic  field attached 
to   stellar cores of  relatively high density, as in the stellar stages from and further the He--burning phase.   We  examine whether  
a field with a scaling as a function of its host core density
is able to exert some significant torque on the outer layers and to produce some coupling of the central region with the above radiative layers and parts of the extended convective envelope. Such a coupling may lead to an acceleration of the rotation  of the outer
layers and a braking of the central core. Among the consequences of this effect,  the contrast of angular velocity  between the core and envelope of the red giants and supergiants may become  smaller, and  the final remnants
may have a slower rotation than predicted by current models which do not account for such effects.

The inclusion of the above effect in models is particularly interesting and topical, since the evolutionary models leading to pulsars always  rotate much too fast
compared to the observed rotation rates of pulsars \citep{Heger2004,Hirschi05}. This is even the case when the magnetic field of the Tayler--Spruit dynamo \citep{Spruit2002}, which could act
in rotating radiative regions,
is accounted for: some internal coupling is introduced, but it is insufficient to slow down the pulsars into the range of the observed rotation rates. %To account for the   very different and high rates of rotation  in the Gamma-Ray-Burst, a different  path with homogeneous evolution during the Main Sequence phase was proposed by\citet{YoonL2005}.
 A similar problem appears for white dwarfs, which generally show rotation velocities  much lower than predicted by standard models \citep{Berger2005}. The inclusion of the magnetic field which could result from the Tayler--Spruit dynamo within radiative regions improves the comparison, but not entirely since a difference of about one order of a magnitude remains for initial masses in the range of 1 to
3  $M_{\odot}$ \citep{Langer2007,Suijs2008}. Here, we must note that the existence of the Tayler--Spruit  dynamo is still strongly debated.
MHD simulations of instabilities in the radiative zone of a differentially rotating star by \citet{ZahnBM2007} well show the presence of  the Pitts and Tayler instability \citep{Pitts1985}, but find no sign of dynamo action. An analysis  with a linear theory of the Tayler instability in the radiative zone of hot stars by \citet{Rudiger2012} also confirms the absence of a dynamo action resulting from
this instability.

Recent asteroseismic observations also  indicate similar trends  about  the too slow rotation of stellar cores in red giants compared to model predictions. The precise 
determinations from  {\it KEPLER} of the rotational frequency splittings of mixed 
modes provide information on the internal rotation of red giants. \citet{Beck2012}
 showed that in the case of the red giant KIC 8366239 the core must rotate about 10 times faster than the surface. 
\citet{Deheuvels2012,Deheuvels2014} analyzed the rotation profiles of six red giants and  found that the cores  rotate between  about 3 to 20  times faster than the envelope. Moreover, they found that the rotation contrast between the core and envelope increases during the subgiant branch, indicating that the cores spin up with time, while the envelopes spin down.
The observed differences of rotation between cores and envelopes are much smaller than predicted by evolutionary models of rotating stars. Analyses with stellar modeling by several  groups  \citep{Eggenberger2012,Marques2013,Ceillier2013,Tayar2013} clearly demonstrate that some
 generally unaccounted physical process is at work producing an important internal viscosity. Thus, there 
is an ensemble of observations pointing in favor of some additional internal coupling in star evolution.

In Sect. \ref{field}, we examine up to which level (the truncation radius) the magnetic field of central regions
may produce some coupling of  a convective envelope. On the basis of standard models of rotating stars, we do some numerical estimates.
In Sect. \ref{rad}, we do the same, but for the intermediate radiative layers subject to horizontal turbulence. In Sect. \ref{exchange},
we examine the timescale of the magnetic coupling and calculate the change of the rotation for the core and for the
external layers. In each case, we make some numerical estimates. Sect. \ref{concl} gives the discussion and conclusions.

\section{The extension of the magnetic braking in a rotating convective 
envelope}\label{field}

%% In a manner similar to \objectname authors can provide links to dataset
%% hosted at participating data centers via the \dataset{} command.  The
%% second curly bracket argument is printed in the text while the first
%% parentheses argument serves as the valid data set identifier.  Large
%% lists of data set are best provided in a table (see Table 3 for an example).
%% Valid data set identifiers should be obtained from the data center that
%% is currently hosting the data.
%%
%% Note that AASTeX interprets everything between the curly braces in the 
%% macro as regular text, so any special characters, e.g. "#" or "_," must be 
%% preceded by a backslash. Otherwise, you will get a LaTeX error when you 
%% compile your manuscript.  Special characters do not 
%% need to be escaped in the optional, square-bracket argument.

%% In this section, we use  the \subsection command to set off
%% a subsection.  \footnote is used to insert a footnote to the text.


%% This section also includes several of the displayed math environments
%% mentioned in the Author Guide.

We consider a rotating stellar core, partly degenerate, of a red giant or red supergiant in the stage of central He burning or in subsequent stages. We assume that a magnetic field of fossil origin is attached to this dense core and that it is extending in the external regions where dissipation occurs. At the surface of the core, the magnetic field has an intensity $B_c$. The cases of a dipolar field is considered in this work. Indeed, a dipolar geometry is  
the preferred one for a magnetic field of fossil origin as demonstrated analytically by \citet{DuezM2010}, who examined the possible magnetic equilibrium configurations of initial fossil fields. This conclusion is in agreement with numerical simulations performed by \citet{BraithwaiteS2004}.

Here, we consider the  intensity $B_c$ of the fossil field in the core as a free parameter of the models, although it is confined within certain limits according to the density scaling as discussed below.
Numerical models will explore the consequences of various field intensities. This may allow us to estimate the range of field intensities necessary 
to reproduce  the  differences of angular velocities between the core and envelope of red giants, as obtained from asteroseismic observations.
 We are guided  in our estimate of the field intensity by the   rough scaling of the pulsar magnetic field  of about $10^{12}$ G  for a density of about $10^{14.6}$ g cm$^{-3}$. Since the pulsar fields are consistent  with the flux conservation of Main Sequence stars \citep{Chanmugam92}, we may hope that this hypothesis is somehow verified in the intermediate stages between Main Sequence and neutron stars. With flux conservation in the
stellar mass during evolutionary stages, we would get for densities of $10^3$, $10^4$, $10^5$, and  $10^6$ g cm$^{-3}$ magnetic fields of $2 \cdot 10^4$,  $9 \cdot 10^4$,  $4 \cdot 10^5$, and   $2 \cdot 10^6$ G
respectively. This is uncertain, but nevertheless indicates the kind of field intensities which could be expected in stellar cores. Thus, we have some guidelines about the range of possible fields in the core of red giants and supergiants, which may be further constrained by asteroseismic observations. In case the new field estimates would be very different, this might reveal other processes, such as efficient dynamos, at work for coupling the cores and envelopes.\\

The field outside the core behaves like 
\begin{equation}
B(r) = B_c \, \left( \frac{R_c}{r} \right)^n \; .
\label{dipole}
\end{equation}

As a result of the flux conservation,  $n=3$ for a dipolar field \citep{Kawaler88,Hartmann98}, (while $n$ would be equal to 2 in a radial spherical case).
%%  However, we shall examine the results of the coupling  for two values of %% n. This may help to test the sensitivity of the results to this parameter. 
We note that for low mass  stars, a value intermediate between $n=2$ and $n=3$ corresponds to the Skumanich law for the long-term  decrease of 
the rotational velocities of solar type stars with time \citep{Skumanich1972}. 
For Ap stars, the observations suggest a variety of geometries, with a dominant large scale dipolar component and strong deviations of axisymmetry, with often some indications of multipolar components \citep{Bagnulo2001} and even strong small scale components \citep{Silvester2011}.
A dipolar field is likely to be preferred on large scales, as mentioned in Sect.1.
Thus, below we will consider the case $n=3$ for the 
field  extending in radiative shells  outside the core and possibly through a part of the extended convective envelope of red giants
or supergiants. For simplification, we assume full magnetic coupling up to a certain truncation radius $r_t$ in the convective envelope and no coupling outside this level. 

This is evidently  a simplified way of treating the complex process of winding up
the field lines and magnetic coupling in the limiting region. Such magnetic interactions between a radiative core and a convective envelope have been studied by several  authors, particularly in the case of the solar spin-down.
By assuming a time independent poloidal field, \citet{Charbonneau1993}
identify  essentially three different phases of the interactions: a rapid buildup
of a toroidal field component in the radiative zone, then phase mixing (interaction of Alfv\'{e}n waves on neighbour field lines) leading to the damping of the large scale toroidal oscillations and coupling of the rotation profile. The third stage is a quasi-stationary evolution with a toroidal field such that its stress, added to the viscous one,
compensates for the  torque at the solar surface, while the rotation rate becomes almost constant on the surfaces of constant poloidal field (Ferraro's state of
isorotation). These three different phases of magnetic  coupling between core and envelope were also well  confirmed by the detailed analysis of \citet{Spada2011}:
-Linear buildup of toroidal field, - Torsional Alfv\'{e}n waves, -Quasi-stationary evolution, where the first two  phases may typically last for less than 1 Myr.
They emphasize the need of a viscosity enhancement by a factor $10^4$ with respect to molecular viscosity and also pointed out that the transport of
angular momentum by the Tayler instability appears negligible. 

 Numerical  3D--MHD simulations of the magnetic coupling between the solar core and envelope were also performed by \citet{Strugarek2011a,Strugarek2011b} with the aim
to examine whether the thinness of the solar tachocline may result from magnetic confinement (this is the transition region from the convective envelope where the rotation varies with latitude to the radiative core where the rotation is almost uniform). They found that the magnetic confinement fails to explain the thinness of the tachocline.
Most interestingly, they demonstrated that
the interior magnetic field does not stay confined in the radiative core, but
expands into the convective zone. This leads  to an efficient outward transport 
of angular momentum through the tachocline, imposing Ferraro's law of iso-rotation over some distance.
However, subsequent numerical models by \citet{Acevedo2013} reconsider the magnetic confinement by a large scale meridional circulation
associated to the convective envelope first proposed by \citet{Gough1998} and support the results of these authors.

\subsection{Estimate of the truncation radius in a convective zone}  \label{trunc1}

The dipolar magnetic field extends in the outer convective zone,  which is subject to vertical motions and strong turbulence,
 which may interfere with  the effects of the magnetic coupling. As, the velocity of the convective motions increases from
the deep interior to reach sonic velocities  in the very external regions with partial hydrogen ionization, the effects associated to convection
increase outwards.  There are several  effects which may influence the penetration of the fossil field in the convective envelope and the extension of the magnetic coupling. The 2--D geometry of the axisymmetric field is important in the global description of the field as shown by \citet{Charbonneau1993}, who also
account for the interaction of the field components with differential rotation and of the damping of magnetic oscillations. Also, we may mention
such effects, as the initial conditions, the Ohmic diffusion, the interaction of the field with the meridional circulations both in the radiative 
and convective regions, the interactions with rotational shears, convective motions and turbulence, an ensemble of effects which were often studied in relation with the problem of  the thinness of the solar tachocline.

 In the present work, we assume that the truncation radius $r_t$ lies at the place 
where the energy density $u_B$ of the magnetic field B is equal to some factor $f$ multiplying the energy density of the convective motions $u_{\mathrm{conv}}$. We introduce this factor $f$ with values  0.1, 1.0 and 10 to account for the  uncertainties associated 
to the multiple 
processes which may influence the magnetic coupling   of a dipolar field to an external convective zone. The use of such a factor  may also allow us to test the robustness of the conclusions we obtain about the coupling.

Below $r_t$, magnetic energy dominates and coupling is present, however the coupling is not instantaneous and in Sect. \ref{exchange} we account for the timescale of the process.
Above $r_t$ convection dominates and there is no magnetic coupling. One has
\begin{equation}
u_B(r) = \frac{B^2(r)}{8 \pi} \quad \quad \mathrm{and} \; \; \; u_{\mathrm{conv}} =\frac{1}{2} \, \varrho \, v^2_{\mathrm{conv}} \; .
\label{ub}
\end{equation}
\noindent
The CGS system of units is used here with 1 Gauss= 1 g$^{1/2}$ cm$^{- 1/2}$ s$^{-1}$.
 \noindent
For the convective velocity, one takes the usual expression appropriate to adiabatic interior stellar regions in the mixing-length theory     
\begin{equation}
v^2_{\mathrm{conv}}= g \, \delta  (\nabla - \nabla_{\mathrm{ad}}) \, \frac{\ell^2}{8 H_P}  \; ,
\end{equation}
\noindent 
where the usual notations of stellar structure have been used. The mixing-length  is usually given by
$\ell= \alpha \, H_P$, where $\alpha\cong \, 1.6$ for the Sun.
The energy density of the convective motions is
\begin{equation}
 u_{\mathrm{conv}} =\frac{1}{16} \alpha^2 \, g \, \varrho \, \delta  (\nabla - \nabla_{\mathrm{ad}}) \, H_P \; .
\end{equation}
\noindent
Expressing the equality
\begin{equation}
u_B(r_t) \, = \,f \; u_{\mathrm{conv}} (r_t), 
\end{equation}
\noindent
we get for the ratio of truncation radius to the core radius in the case of a dipolar field,
\begin{equation}
\left( \frac{r_t}{R_c} \right) \, = \, \left( \frac{2 \, B^2_c}{f \; \pi \alpha^2 \, g \varrho \delta (\nabla - \nabla_{\mathrm{ad}}) H_P} \right)^{\frac{1}{6}}  \; .
\label{rtrc}
\end{equation}
\noindent
We notice the very weak dependence on the factor $f$ and also see that the truncation radius in the convective envelope increases slowly with the magnetic field of the core; for lower densities and gravities, it extends further out. 

\subsection{Numerical estimates during the  red supergiants} \label{num1}

%\begin{figure}[t]
%\begin{center}
%\includegraphics[width=9.0cm, height=6.5cm]{schema.eps}
%\caption{Schematic representation of the various levels considered
%in the model of magnetic braking: the core in red, the radiative shell
%burning layers in yellow, the convective zone in pink.}
%\label{schema}
%\end{center}
%\end{figure}

Let us  consider the case of  a model of a 15 $M_{\odot}$ star with solar metallicity $Z=0.014$ and an initial rotation velocity equal to 40\%  of the critical value \citep{Ekstrom12}.  We first examine the situation during the helium burning phase when the central helium mass fraction is
$Y_c =0.30$  (age = $14.467535 \cdot 10^6$ yr.), with a central density   of $1.26 \cdot 10^3$ g cm$^{-3}$. %  while the average core density is  $1.35 \cdot 10^2$ g cm$^{-3}$. 
The star with an actual mass of 12.991 $M_{\odot}$ is in the stage of red supergiant with $\log (L/L_{\odot})=4.868$ and 
$\log  T_{\mathrm{eff}}=3.566$.
The core totally deprived of hydrogen ($X_c < 10^{-10}$)  extends up to $M_r/M=0.328$ with a radius of $r=2.164 \cdot 10^{10}$ cm,
while the base of the convective envelope lies at $M_r/M=0.687$, with  $r=3.954 \cdot 10^{12}$ cm, i.e. at  a distance equal to
182.7 times the size of the core. At such a  large distance, the dipolar field has so much decreased that it is not likely to exert some coupling of the convective envelope. Only an extreme dipolar core field, with a very unrealistic intensity of $10^{11}$ G or more could build an efficient coupling of the convective envelope at this stage, while  a field   of about $2.2 \cdot 10^4$ G would be expected according to 
the central density and the pulsar scaling.

The same mass model at the end of the central C-burning stage at an age of $15.065866 \cdot 10^6$ yr.  with a central density  
 of $4.37 \cdot 10^6$ g cm$^{-3}$ looks slightly less unfavorable.  The core totally deprived of hydrogen   extends further out 
up to  $M_r/M=0.4260$ with a radius of $r= 2.357 \cdot 10^{10}$ cm,
while the base of the convective envelope goes deeper than during the He-burning phase  down to $M_r/M=0.469$, with $r=2.565 \cdot 10^{11}$ cm, i.e. at  a distance equal to only 10.88 times the size of the core. The estimate of the ratio $({r_t}/{R_c} )$ from (\ref{rtrc})
in the region at the base of the convective envelope indicates that, due to the fast decrease of the dipolar field, a field in the core of  $2.5 \cdot 10^8$ G  (for $f=1$)  would be necessary  to produce the coupling of the core up to the base of the convective envelope. For $f$-values of 0.1 or 10,
the necessary core field would be   $8 \cdot 10^7$ G  and  $8 \cdot 10^8$ G respectively.
Now, adopting the rough scaling of pulsar fields mentioned in Sect. \ref{field}, we see that the central density
  would imply a magnetic field of $5 \cdot 10^6$ G. Thus, we see that, even if  $f$ is equal to 0.1, it seems very unlikely that a coupling of the regions at  the base of the convective envelope is possible, and this by a relatively wide margin.

We note that the envelope is extremely extended, the star radius is 886.2 $R_{\odot}$, which represents 2625 times the core radius. Near the stellar surface, the magnetic dipolar field would be as low as  $8.4 \cdot 10^{-4} $ G and totally unable to produce some coupling of the
outer layers subject to nearly sonic velocities. We note that this value is very different from the value  of a few Gauss  found at the
surface of red giants  \citep{Konstantinova-Antova2013}.  We take this as an indication that  the fossil field does not permeate the huge convective envelope, in which the
interaction of convective motions and rotation may produce some dynamo like in the solar case. The observed field  may thus  be a result of 
the stellar convective dynamo, rather than a signature of  central fossil field. 

 From these numerical estimates, we  reach the following conclusion.
Magnetic fields  consistent with  the pulsar field-density scaling (Sect. \ref{field}),  in the core of stars in the He--burning or later evolutionary phases, appear unable 
to produce a magnetic coupling of the large convective envelopes present in red giants and supergiants.

\section{The truncation radius in a rotating radiative zone} \label{rad}

We have seen above  that in the stages of helium burning, there may be a large   zone  between the core and the outer convective envelope. The intermediate region may be totally or partly occupied by a radiative zone. Indeed,
depending on mass loss, there may exist some small convective zones in this intermediate region. If so, within such convective zones, the appropriate expression (\ref{rtrc}) of the truncation radius also applies.  

 In the intermediate radiative region, many physical variables are generally changing rapidly, such as the density, the composition and also the angular velocity $\Omega$. The layers in differential rotation are subject to turbulence both vertical and horizontal.
The horizontal turbulence is the largest, since it is not damped by the vertical density stratification \citep{Zahn92}. This  turbulence is characterized by a 
diffusion coefficient $D_{\mathrm{h}}$, which is rather uncertain and various expressions have been proposed for it
\citep{Zahn92,Maeder2003, Mathis2004}. Moreover, the horizontal turbulence can be modified by the presence of the field.

\subsection{Estimate of the truncation radius in a horizontal turbulent zone}

The velocity $v_{\mathrm{h}}$ of the horizontal turbulence can be written
\begin{equation}
v_{\mathrm{h}} \, = \, \frac{D_{\mathrm{h}}}{r} \; ,
\end{equation}
\noindent 
since the turbulence is highly anisotropic there is  no factor $1/3$ here. The corresponding energy density $u_{\mathrm{h}}$  is
\begin{equation}
u_{\mathrm{h}} \, = \, \frac{1}{2} \, \varrho \, \, \frac{D^2_{\mathrm{h}}}{r^2} \; .
\end{equation}
\noindent
As in Sect. \ref{trunc1}, we define the truncation by the location where the energy densities are equal, also considering an $f$ factor
equal to 0.1 or 10 in order to account the many uncertainties intervening in the coupling,
\begin{equation}
u_B(r_t) \, = \, f \,  u_{\mathrm{h}} (r_t)  \, . 
\end{equation}
With  (\ref{dipole}) and (\ref{ub}), this gives
\begin{equation}
\frac{B^2_{\mathrm{c}}}{4 \pi} \, \frac{R^{6}_{\mathrm{c}}}{r_t^{4}} \, = \,f \, \varrho \, D^2_{\mathrm{h}} \; .
\end{equation}
\noindent
We obtain for the ratio of the truncation to the core radius,
\begin{equation}
\left( \frac{r_t}{R_c} \right) \, = \, \left( \frac{ B^2_c \, R^2_c}{4 \pi \,f \,  \varrho \, 
D^2_{\mathrm{h}}} \right)^{\frac{1}{4}} \; .
\label{rtrcdh}
\end{equation}
\noindent
Consistently, this ratio increases with the magnetic field and decreases with increasing turbulence.
We see that the truncation  is more sensitive to the field intensity in the case of horizontal turbulence than in the case of convective
motions. We also notice that the truncation radius depends very weakly on the power  $-(1/2)$ of the coefficient of horizontal turbulence.

The expression of  the coefficent of horizontal turbulence $D_{\mathrm{h}}$ currently used in Geneva models in that by
\citet{Zahn92},
\begin{equation}
D_{\mathrm{h}} \,=  \, \frac{1}{c_{\mathrm{h}}} \, r \left|2\,  V_2 - \alpha \, U_2\right| \; ,
\end{equation}
where $V_2$ and $U_2$ are respectively the amplitudes of the radial and horizontal components of the velocity of
meridional circulation. The coefficient $c_{\mathrm{h}}$ is normally $\leq 1$, but is generally taken equal to 1, while
$\alpha = (1/2) [dln(r^2 \,\Omega)/ dlnr]$, which is equal to 1 in region of uniform rotation. Below, we mention the cases of the other 
diffusion coefficients  of horizontal turbulence  \citep{Mathis2004,maederlivre09}.

\subsection{Numerical estimates in radiative zones of red supergiants} \label{num2}

We first consider  the same model as in Sect. \ref{num1} of a 15 $M_{\odot}$  star at the stage of central He--burning  with $Y_{\mathrm{c}}=0.30 $  in the red supergiant phase. We recall that in this model the base of the convective envelope lies at a distance 
equal to 182.7 times the size of the convective core. The layers between the core and the convective envelope are fully radiative.
(In some cases, there could be some intermediate convective zones, particularly in models with little
or no mass loss, but in the present model the whole intermediate zone is  radiative).

The horizontal turbulence is $D_{\mathrm{h}}= 8.11 \cdot 10^{10}$ cm$^2$ s$^{-1}$ and the  density $\varrho = 3.85 \cdot 10^{-6}$ near the top of  the intermediate zone (at $M_r/M =0.673 $, this means at 1.4 \%  in mass fraction below the bottom of the outer convective zone). We make such a choice in order that the value of $D_{\mathrm{h}}$ is not influenced by some edge effects. 
We find with these values and expression (\ref{rtrcdh})
  that a dipolar field in the core equal to about $ 8.7 \cdot  10^2$ G 
would be sufficient   to  fully couple the intermediate zone to the core. Such core fields are much lower than  the one approximately suggested
 ($2.2 \cdot 10^4$ G) by the pulsar field scaling on the actual central core density ($1.26 \cdot 10^3$ g cm$^{-3}$), as given at the beginning of Sect. \ref{field}. Thus, we conclude that the intermediate radiative zone is likely to be fully magnetically coupled to the core. This conclusion applies whatever the
factor $f$, since we consider the top of the intermediate radiative zone. From the
dependence of $(r_t/R_c)$ on $B_c$ and $D_h$ in expression  (\ref{rtrcdh}), we notice that the above
conclusion on the coupling remains valid even if $D_h$ is larger by one order of
magnitude than given here. However, this would not necessarily be the case if
$D_h$ would be larger by several orders or magnitude and become close to values
of diffusion more typical of convective zones.

For the model of Sect. \ref{num1} at the end of the C--burning phase, the intermediate radiative  zone extends up to 10.88 times the radius of the core. 
The coefficient of horizontal turbulence ($D_{\mathrm{h}}= 4.69 \cdot 10^8$ cm$^2$ s$^{-1}$) near the top of the radiative zone  is much smaller than the coefficient of 
convective diffusion of the order of a few $10^{16}$ cm$^2$ s$^{-1}$ at the base of the convective zone. This means that the perturbations of the  magnetic field by the horizontal turbulence in the radiative region
are extremely small compared to the effects of the turbulence in the  convective zone. We find that even a dipolar field of 1 G  in the core would be sufficient to couple the radiative zone up to its top.

 We  conclude the following:

- Radiative rotating stellar zones, subject to  horizontal turbulence, are much more easily coupled to the core than convective regions. The reason is 
the much lower turbulent diffusion coefficient than the one characterizing convection.

- Even weak dipolar fields attached to the core are sufficient to couple an extended
intermediate radiative zone up to the base of the convective envelope. This applies to the phase of He-burning and 
is even more likely in further evolutionary phases.  
 However, the overall results also 
depend on the timescales of the process as shown in the next Section.

\section{The exchange of angular momentum}\label{exchange}

The magnetic coupling between the core and the external layers produces a slowing down of the core 
and a gain  of angular momentum by the outer layers. The treatment of the problem is also depending whether these layers are 
radiative or convective.

\subsection{The reduction of the angular momentum in the core}

Let us examine the reduction of the angular momentum $J_c$ of the core of mass $M_c$, radius $R_c$  and angular velocity $\Omega_c$ due to the coupling with the above layers up  to truncation radius $r_t$. Assuming that the spherical symmetry of the inner layers is not much modified by rotation, we write
the angular momentum of the core, 
\begin{equation}
J_c= \left(\frac{2}{3}\right) \Omega_c  \int^{M_c}_0 r^2 dM_r       \; .
\end{equation}
\noindent
The core, whether convective or radiative, is  supposed to rotate uniformly due to the internal magnetic coupling.
The reduction of the angular momentum of the core due to the coupling of the layers from $M_c$ to $M(r_t)$ is
\begin{eqnarray}
\delta J_c = %- \left(\frac{2}{3} \right) \,4  \pi \int^{r_t}_{R_c}  r^4 \, \varrho(r) \, [\Omega_c -\Omega(r)]   \, dr = 
-  \left( \frac{2}{3} \right) \, \int^{M(r_t)}_{M_c}  r^2 \,  [\Omega_c -\Omega(M_r)]   \, dM_r \, .
\end{eqnarray}
\noindent
since the core generally  rotates faster than the envelope, as a result of core contraction.
The rate $(dJ_c/dt)$ of the  change of the angular momentum of the core provides
the change of the angular velocity of the core.
\begin{equation}
\frac{1}{\Omega_c} \, \frac{d\Omega_c}{dt} \, = \, \frac{1}{J_c} \,\frac{dJ_c}{dt} \; .
\end{equation}
\noindent
Here, we must not account for the change of the momentum of inertia of the core  over a time step,
because the consequences of  stellar expansion or contraction on the transport of angular momentum
are already accounted for in the general equation describing  the evolution of $\Omega(M_r)$, see for 
example expressions (10.121) and (10.122) in \citet{maederlivre09}.
 
The frequency of a magnetic wave is the Alfv\'{e}n frequency $\omega_A$,
\begin{equation}
\omega_A (M_r) \, = \, \frac{ B(M_r)}{r \, (4 \pi \, \varrho)^{1/2}} \; .
\end{equation}
\noindent
The characteristic timescale for the exchange of angular momentum between differentially rotating layers
is the Alfv\'en  timescale $\omega^{-1}_A(M_r)$, calculated with the poloidal component of the magnetic field \citep{Mestel70}.
The rate of change of the angular momentum thus becomes
\begin{equation}
\frac{d J_c}{dt} = -  \left( \frac{2}{3} \right) \,  \int^{M(r_t)}_{M_c}  r^2 \,  [\Omega_c -\Omega(M_r)]  \, \omega_A (M_r) \, \, dM_r \, .
\label{djdt}
\end{equation}
\noindent
At each time step, this expression permits (with the parameters of the considered time step) to calculate the change of the rotation of the regions within the radius
$R_{\mathrm{c}}$. Some attention is to be paid to the definition of the core of mass $M_c$. It is not necessarily the mass of the  convective
core at the considered evolutionary phase, since in some advanced nuclear phases central convection is even absent. We suggest to 
consider that the outer limit of the  
core with strong magnetic field is located in the region of the very steep density gradient, which separates the central region deprived of
hydrogen  from the envelope where hydrogen is present. The variation of the magnetic field with the density at the power 2/3 applies when the mass and the magnetic flux remains constant.
After the end of core He-burning, the core mass has essentially reached its final value, but
its radius can change, so that from this point onwards the dependence of the field on density is
likely very well represented by the 2/3 exponent.
Detailed numerical models will allows us to explore the domain of parameters.

\subsection{Gain of angular momentum by the external  layers}

Now, we have also to be concerned by the gains of angular momentum and changes of angular velocity  of the magnetically coupled 
layers external to the core. Let us first examine the case of the radiative layers above the core. Each layer of mass thickness 
$\Delta M_r$  (the width of the mass steps in the models) between radius $r_1$ and $r_2$   ($r_2 > r_1$) has a certain momentum of inertia $\Delta I (M_r)$, at the time considered,
\begin{equation} 
\Delta I(M_r) \, = \, \frac{2}{5} \, \, \frac{r^5_2 - r^5_1}{r^3_2 - r^3_1} \, \Delta M_r \, .
\end{equation}
\noindent
A mass shell is usually characterized by a mean value of the density and thus this assumption applies.
The shell has a certain amount of angular momentum $\Delta J(M_r)$,
\begin{equation}
\Delta J(M_r) \, = \, \Delta I(M_r) \, \Omega(M_r) \, .
\label{dj}
\end{equation}
\noindent
The rate of increase  of the angular momentum due to the magnetic torque  on this shell per unit of time  is,
\begin{equation}
\frac{d [\Delta J(M_r)]}{d t} =  \Delta I(M_r)  \,  [\Omega_c -\Omega(M_r)]  \, \omega_A (M_r)  \, .
\label{djc}
\end{equation}
\noindent
Thus, the corresponding rate of change of the angular velocity  $\Omega (M_r)$ is
\begin{equation}
\frac{1}{\Omega(M_r)} \, \frac{d\Omega (M_r)}{dt} \, = \, \frac{1}{\Delta J(M_r)} \,\frac{d [\Delta J(M_r)]}{dt} \; .
\label{io}
\end{equation}
\noindent
Again,  we have not to account for the change of the momentum of inertia for the same reasons as above.  Expression (\ref{io}) can be rewritten with the help of (\ref{djc}) and (\ref{dj}) at the level $M_r$ (which we now generally omit in the equations below),
\begin{eqnarray}
 \frac{d\Omega }{dt}  = \Omega  \frac{1} {\Delta J}\Delta I  \,  [\Omega_c -\Omega]  \, \omega_A   =
[\Omega_c -\Omega]  \, \omega_A  \; .
\end{eqnarray}
\noindent
This just means that for a shell  of a given mass and momentum of inertia the change of $\Omega$ due to the magnetic torque  is   equal to   the difference
of angular velocity between the  core and the considered layer multiplied by the ratio of the interval of time $dt$ to  $(1/ \omega_A )$, the characteristic time of the coupling. This differential equation is easily integrated, if we consider that $\omega_A$ is constant, which is valid over a small interval of time $\Delta t$.
The solution is 
\begin{eqnarray}
[\Omega_c - \Omega(\Delta t)] \, = \, [\Omega_c - \Omega(\Delta t)]_0 \, e^{-(\omega_A \, \Delta t)}  \; ,
\end{eqnarray}

\noindent
at the mass level $M_r$. It gives the $\Omega$--difference after the time interval $\Delta t$ as a function of the 
$\Omega$--difference 
at the beginning of this time interval. Since $\omega_A$ is not a constant, but a slowly varying function of time, the interval
$\Delta t$ must at each level be small with respect to $(1/ \omega_A )$. In case the timescale $(1/ \omega_A )$ at a certain level $M_r$ would be very small,
i.e. negligible with respect to the evolutionary timescale, the solution is simply to assume $\Omega(M_r) = \Omega_c$, i.e. that magnetic equilibrium is instantaneous.
The effects of meridional circulation and shears, when present,  have also to be included, as usual in models of rotating stars, for the 
calculation of the evolution of $\Omega(M_r)$.\\   % Let us note that, in the integration of the equations of stellar structure, the variablesdepending on $M_r$ have to be estimated in the middle of the interval $\left[ M_r, M_r + \Delta M_r \right]$ for a proper integration of the structure equations. \\

For a convective envelope, some further operations are  needed. Such envelopes are often considered, to the first order,  as having a solid body rotation, due to the 
very high viscosity associated to convective turbulence.  In such case, one  also has to estimate by (\ref{djc}) the rates of change  $(d [\Delta J(M_r)]/{d t})_i$ of the angular momentum
for all  convective layers $i=1$ to $N$ up to the truncation radius.  In this way, the various changes $\Delta [\Delta J(M_r)]_i$ over 
a time step $\Delta t$ are obtained.  Then, by summing up
 all the changes  one gets the total change  $\Delta J_{\mathrm{conv}}$ of angular momentum of the convective envelope due to magnetic coupling during that interval of time,
\begin{equation}
\Delta J_{\mathrm{conv}} = \sum_{i=1,N} \Delta [\Delta J(M_r)]_i  \; .
\end{equation}
\noindent
Now, the change $\Delta \Omega_{\mathrm{conv}}$ of angular velocity  $\Omega_{\mathrm{conv}}$ of the convective envelope
over the time interval $\Delta t$ is obtained by 
\begin{equation}
\frac{\Delta \Omega_{\mathrm{conv}}}{\Omega_{\mathrm{conv}}} \, = \, \frac{\Delta J_{\mathrm{conv}}}{ J_{\mathrm{conv}}}  \; .
\label{jom}
\end{equation}
\noindent
The usual treatment of convective envelopes in our models conserves the angular momentum of the convective envelope with a proper account of the change of the momentum of inertia. Thus,  Eq.~(\ref{jom}) represents only the change  of angular velocity due to the magnetic torque.

There the angular momentum  $J_{\mathrm{conv}}$ is given by
\begin{equation} 
J_{\mathrm{conv}} \, = \left( \frac{2}{3} \right) \, \Omega_{\mathrm{conv}} \, \int^{M_{\mathrm{top}}}_{M_{\mathrm{base}}}  r^2 \, dM_r \, ,
\end{equation}
\noindent
where the integration is performed from the base to the top of the convective envelope. In red giants and supergiants, the top of this envelope is the total stellar radius.

\subsection{Numerical estimates of the timescale of the coupling during the helium burning phase}  \label{numtime1}

The timescale of the magnetic coupling is a major characteristic of the process determining the kind of distribution of $\Omega$  in the stars at the various phases.
Let us first consider  the same model of a 15 $M_{\odot}$ in the stage of central helium burning with $Y_c = 0.30$ 
as in Sect. \ref{num1} and \ref{num2}.  We have seen that at this stage the magnetic coupling is possible for the intermediate zone, while it seems 
unlikely for the outer convective envelope.
At the edge of the core deprived of hydrogen ($Mr/M = 0.328$),  the Alfv\'en frequency for a typical field in the core  of   $2.2 \cdot 10^4$ G  as given by the pulsar scaling is $\omega_{\mathrm{A}}=
4.96 \cdot 10^{-8}$ s$^{-1}$. This  corresponds to a timescale of 0.64 yr. 
% (The angular velocity at the edge of the core is $\Omega = 1.56 \cdot 10^{-4}$ s$^{-1}$, corresponding to a rotation period $2 \pi/\Omega = 11.2 $ h.
 Thus, in view of the long timescale
of the He--burning phase ($1.49  \cdot 10^6$ yr), we conclude that the coupling is extremely fast at the edge of the core.

Now, we examine the timescales near the outer edge  of the intermediate zone, at $M_r/M= 0.673$ like in Sect. \ref{num2}. There, for the same typical intensity of the dipolar field  in the core
of $2.2 \cdot 10^4$ G, we have $\omega_{\mathrm{A}}=
4.07 \cdot 10^{-13}$ s$^{-1}$, corresponding to an Alfv\'en timescale of $7.8 \cdot 10^4$ yr. %In the radial case,  $\omega_{\mathrm{A}}=5.61 \cdot 10^{-11}$ s$^{-1}$   and the timescale is  565 yr.
%The local angular velocity is  $\Omega = 3.24 \cdot 10^{-8}$ s $^{-1}$, corresponding to a rotation period of 6.1 yr.
%The characteristic time of the magnetic equilibrium is $\Omega/(\omega_{\mathrm{A}})^2 = 6.1\cdot 10^{9}$ yr in the dipolar case  and  $3.3 \cdot 10^5$ yr in the radial case.
 This timescale is  longer than the one  at the edge of the core. 
This value is shorter than the timescale  of $1.49  \cdot 10^6$ yr of the  He-burning phase of  a rotating star of 15 $M_{\odot}$. Thus, for fields  in the range of $10^3$  to $10^5$ G, compatible with the pulsar scaling (Sect. \ref{field}),
the timescale for the magnetic coupling of the dipolar field  is short enough that we may expect full
coupling up to the truncation radius. We recall that this radius is  unlikely to be in the outer convective envelope, but the intermediate radiative zone is likely fully coupled.

%- The Kelvin--Helmholtz timescale, characterizing the transition from the H- to the He-burning phase is of the order of $10^5$ yr for a rotating star of  15 $M_{\odot}$. Thus, during the contraction phase between the H-- and He--burning phase, the coupling of the intermediate zone may beonly partial in the dipolar case, thus allowing some significant $\Omega$--gradient between the core and envelope, while in the radial case the coupling would be complete.

Globally, we can say that the red giants and supergiants in the He--burning phase will experience no direct magnetic coupling
between the core and the outer convective envelope, but the intermediate radiative zone between the core and the envelope
will be magnetically coupled to the core. This intermediate radiative zone at the stage  $Y_c =0.30$ contains a fraction of the
stellar mass comparable to that of the core deprived of hydrogen, while the  momentum of inertia of the intermediate zone is
at least one order of magnitude larger than that 
of the core. Thus, one may expect a strong braking of the core rotation during the He--burning phase.  
Through the other transport  mechanisms, like shear, meridional circulation and convection,
some transport of angular momentum to the outer convective envelope may also result.

\subsection{The timescale at the end of the phase of central C--burning}  \label{numtime2}

 We now
examine the timescale at the edge of the core  deprived of hydrogen ($M_r/M= 0.426$). % The surface of the CO--core is at $M_r/M = 0.261$). 
 We consider 
a field of $5 \cdot 10^6$ G at the surface of the core, consistent with the pulsar scaling.  The Alfv\'en frequency is $1.90 \cdot 10^{-5}$ s$^{-1}$, which corresponds to a timescale of 14.6 h. %and the angular velocity $2.536 \cdot 10^{-4}$ s$^{-1}$. The two frequencies differ by only one order of a magnitude. 
The  coupling is thus instantaneous at the edge of the core.

We have seen that the intermediate zone may experience the magnetic torque, but not the convective envelope.
 Thus, we  consider the situation  near the top of the intermediate radiative zone as before. The same dipolar field of $5 \cdot 10^6$ G at the surface of the  core
 leads here to a field $4.41 \cdot 10^3$ G.
The corresponding Alfv\'en frequencies are $2.97 \cdot 10^{-7}$ s$^{-1}$, which corresponds to a  timescale of 0.11 yr. 
%The angular velocity is $\Omega=2.465 \cdot 10^{-7}$, which is of the same order or lower than  the Alfvén frequency, so that  the Coriolis effects play only a little role.
Thus, the coupling of the core 
with the entire intermediate radiative zone is very fast   in the C-burning phase and  full coupling up to the truncation radius is expected. 
%We have seen that the  magnetic coupling from the core may marginally reach of the  innermost layers of the convective envelope.
 
\section{Conclusions} \label{concl}

These are first estimates of the possible  magnetic  coupling between dense stellar
cores and external layers based on a simple modelling. Despite this fact,
 the  first estimates of the order of magnitude of the main intervening effects
support   some  clear conclusions:

-  The direct magnetic coupling of the convective envelopes with the stellar cores is very 
unlikely throughout the advanced evolution.

-  However, the magnetic coupling between the core and the intermediate radiative 
zone appears very effective during the He-burning phase in red supergiants and 
even more in later phases. 

- This coupling of the core and intermediate radiative zone  is able to produce a  strong spin down 
of the core of red giants and supergiants, leading also to relatively  slowly rotating stellar remnants, like 
white dwarfs and pulsars.

- The slowly rotating  envelopes of supergiants will  receive some additional angular momentum
 from the He-burning phase and later, for two reasons. First, because the other processes 
of angular momentum transport, like shear and circulation, transmit anyway some angular 
momentum between the radiative and convective envelope and secondly because as
 evolution proceeds the convective envelope gets much deeper and encompasses layers 
which have been spun up previously.

As mentioned in the introduction, magnetic braking may intervene at all
stages of stellar evolution. The  properties of
the magnetic field needs to be further studied by detailed numerical simulations  through the various evolutionary stages.
The first results presented here, which seem rather robust in view of the orders of magnitude obtained, are encouraging for future study concerning the magnetic coupling between stellar 
cores and the adjacent radiative layers. 

\noindent
Acknowledgments: We express our best thanks to Dr. S. Mathis for  his many very appropriate and helpful remarks.


\begin{thebibliography}{22}
\expandafter\ifx\csname natexlab\endcsname\relax\def\natexlab#1{#1}\fi


\bibitem[{{Acevedo-Arreguin}  {et al.}(2013)}]{Acevedo2013}
{Acevedo-Arreguin}, L.A., {Garaud}, P.  \& {Wood}, T.S.  2013, \mnras, 434, 720


\bibitem[{{Bagnulo}(2001)}]{Bagnulo2001}
{Bagnulo}, S. 2001, in Astronomical Society of the Pacific Conference Series,
  Vol. 248, Magnetic Fields Across the Hertzsprung-Russell Diagram, ed.
  G.~{Mathys}, S.~K. {Solanki}, \& D.~T. {Wickramasinghe}, 287

\bibitem[{{Beck} {et~al.}(2012){Beck}, {Montalban}, {Kallinger}, {De Ridder},
  {Aerts}, {Garc{\'{\i}}a}, {Hekker}, {Dupret}, {Mosser}, {Eggenberger},
  {Stello}, {Elsworth}, {Frandsen}, {Carrier}, {Hillen}, {Gruberbauer},
  {Christensen-Dalsgaard}, {Miglio}, {Valentini}, {Bedding}, {Kjeldsen},
  {Girouard}, {Hall}, \& {Ibrahim}}]{Beck2012}
{Beck}, P.~G., {Montalban}, J., {Kallinger}, T., {et~al.} 2012, \nat, 481, 55

\bibitem[{{Berger} {et~al.}(2005){Berger}, {Koester}, {Napiwotzki}, {Reid}, \&
  {Zuckerman}}]{Berger2005}
{Berger}, L., {Koester}, D., {Napiwotzki}, R., {Reid}, I.~N., \& {Zuckerman},
  B. 2005, \aap, 444, 565


\bibitem[{{Braithwaite} \& {Spruit}(2004)}]{BraithwaiteS2004}
{Braithwaite}, J. \& {Spruit}, H.C. 2004, Nature, 431, 819


\bibitem[{{Ceillier}  {et al.}(2013)}]{Ceillier2013}
{Ceillier}, T., {Eggenberger}, P., {Garcia}, R.A. \& {Mathis}, S.,
   2013, \aap, 555, 54

\bibitem[{{Chanmugam}(1992)}]{Chanmugam92}
{Chanmugam}, G. 1992, \araa, 30, 143


\bibitem[{{Charbonneau} \& {MacGregor}(1993)}]{Charbonneau1993}
{Charbonneau}, P. \& {MacGregor}, K.B. 1993, \apj, 417,762


\bibitem[{{Deheuvels}  {et al.}(2012)}]{Deheuvels2012}
{Deheuvels}, S.,{Garcia}, R.A., {Chaplin}, W.G., {Basu}, S.,
 {Antia}, H.M.  {et al.},  2012, \apj, 756, 19


\bibitem[{{Deheuvels}  {et al.}(2014)}]{Deheuvels2014}
{Deheuvels}, S.,{Dogan}, G., {Goupil}, M.J., {Appourchaux}, T.,
 {Benomar}, O.  {et al.},  2014, \aap, 564, 27


\bibitem[{{Duez} \& {Mathis}(2010)}]{DuezM2010}
{Duez}, V. \& {Mathis}, S. 2010, \aap, 517, 58


\bibitem[Duez et al.(2010)]{Duez2010} Duez, V., Braithwaite, J., 
\& Mathis, S.\ 2010, \apjl, 724, L34 

\bibitem[{{Eggenberger} {et~al.}(2012){Eggenberger}, {Montalb{\'a}n}, \&
  {Miglio}}]{Eggenberger2012}
{Eggenberger}, P., {Montalb{\'a}n}, J., \& {Miglio}, A. 2012, \aap, 544, L4

\bibitem[{{Ekstr{\"o}m} {et~al.}(2012){Ekstr{\"o}m}, {Georgy}, {Eggenberger},
  {Meynet}, {Mowlavi}, {Wyttenbach}, {Granada}, {Decressin}, {Hirschi},
  {Frischknecht}, {Charbonnel}, \& {Maeder}}]{Ekstrom12}
{Ekstr{\"o}m}, S., {Georgy}, C., {Eggenberger}, P., {et~al.} 2012, \aap, 537, A146


\bibitem[{{Gough} \& {McIntyre}(1998)}]{Gough1998}
{Gough}, D. \& {McIntyre}, M.E. 1998, Nature,  394, 755



\bibitem[{{Hartmann}(1998)}]{Hartmann98}
{Hartmann}, L. 1998, {Accretion Processes in Star Formation}, ed. {Cambridge University Press}

\bibitem[{{Heger} {et~al.}(2004){Heger}, {Woosley}, {Langer}, \&
  {Spruit}}]{Heger2004}
{Heger}, A., {Woosley}, S.~E., {Langer}, N., \& {Spruit}, H.~C. 2004, in IAU
  Symposium, Vol. 215, Stellar Rotation, ed. A.~{Maeder} \& P.~{Eenens}, 591

\bibitem[{{Heger} {et~al.}(2005){Heger}, {Woosley}, \& {Spruit}}]{Heger2005}
{Heger}, A., {Woosley}, S.~E., \& {Spruit}, H.~C. 2005, \apj, 626, 350

\bibitem[{{Hirschi} {et~al.}(2005){Hirschi}, {Meynet}, \& {Maeder}}]{Hirschi05}
{Hirschi}, R., {Meynet}, G., \& {Maeder}, A. 2005, \aap, 443, 581

\bibitem[{{Kawaler}(1988)}]{Kawaler88}
{Kawaler}, S.~D. 1988, \apj, 333, 236


\bibitem[{{Konstantinova-Antova} {et~al.}(2013){Konstantinova-Antova}, {Auriere}, {Charbonnel}, {Drake}, {Wade} \& {et al.}}]{Konstantinova-Antova2013}
{Konstantinova-Antova}, R., {Auri\`{e}re}, M.,  {Charbonnel}, C., {Drake}, N., {Wade}, G. \& {et al.}, A. 2013, arXiv: 1311.0482


\bibitem[{{Landstreet} {et~al.}(2012){Landstreet}, {Bagnulo}, {Valyavin},
  {Fossati}, {Jordan}, {Monin}, \& {Wade}}]{Landstreet2012}
{Landstreet}, J.~D., {Bagnulo}, S., {Valyavin}, G.~G., {et~al.} 2012, \aap,
  545, A30

\bibitem[{{Langer}(2007)}]{Langer2007}
{Langer}, N. 2007, in Astronomical Society of the Pacific Conference Series,
  Vol. 372, 15th European Workshop on White Dwarfs, ed. R.~{Napiwotzki} \&
  M.~R. {Burleigh}, 3


\bibitem[{{Maeder} (2003)}]{Maeder2003}
{Maeder}, A.  2003, \aap, 399, 263



\bibitem[{{Maeder}(2009)}]{maederlivre09}
{Maeder}, A. 2009, {Physics, Formation and Evolution of Rotating Stars}, ed.
  {Springer Berlin Heidelberg.}

\bibitem[{{Maeder} \& {Meynet}(2003)}]{magn03}
{Maeder}, A. \& {Meynet}, G. 2003, \aap, 411, 543

\bibitem[Markey 
\& Tayler(1973)]{Markey73} Markey, P., \& Tayler, R.~J.\ 1973, \mnras, 163, 77 


\bibitem[{{Marques}  {et al.}(2013)}]{Marques2013}
{Marques}, J.P.,{Goupil}, M.J., {Lebreton}, Y., {Talon}, S.,
 {Palacios}, A.  {et al.},  2013, \aap, 549, A74


\bibitem[{{Mathis}  {et al.}(2004)}]{Mathis2004}
{Mathis}, S., {Palacios}, A. \& {Zahn}, J.-P. 2004, \aap, 425, 243



\bibitem[{{Mathis} \& {Zahn}(2005)}]{MathisZ2005}
{Mathis}, S. \& {Zahn}, J.-P. 2005, \aap, 440, 653


\bibitem[{{Mestel}(1970)}]{Mestel70}
{Mestel}, L. 1970, M\'{e}moires of the Soci\'{e}t\'{e} Royale des Sciences de Li\`{e}ge, 19,
  167

\bibitem[{{Meynet} {et~al.}(2011){Meynet}, {Eggenberger}, \&
  {Maeder}}]{MEM2011}
{Meynet}, G., {Eggenberger}, P., \& {Maeder}, A. 2011, \aap, 525, L11


\bibitem[{{Pitts} \& {Tayler}(1985)}]{Pitts1985}
{Pitts}, E.,  {Tayler}, R.J. 1985, \mnras, 216, 139



\bibitem[{{R\"{u}diger} {et~al.}(2012){Rudiger}, {Kitchatinov}, \&
  {Elstner}}]{Rudiger2012}
{R\"{u}diger}, G., {Kitchatinov}, L.L., \& {Elstner}, D. 2012, \mnras, 425, 2267

\bibitem[{{Silvester} {et~al.}(2011){Silvester}, {Wade}, {Kochukhov},
  {Landstreet}, \& {Bagnulo}}]{Silvester2011}
{Silvester}, J., {Wade}, G., {Kochukhov}, O., {Landstreet}, J., \& {Bagnulo},
  S. 2011, in Astronomical Society of the Pacific Conference Series, Vol. 449,
  Astronomical Society of the Pacific Conference Series, ed. P.~{Bastien},
  N.~{Manset}, D.~P. {Clemens}, \& N.~{St-Louis}, 280

\bibitem[{{Skumanich}(1972)}]{Skumanich1972}
{Skumanich}, A. 1972, \apj, 171, 565


\bibitem[Spada et al.(2011)]{Spada2011} Spada, F., Lanzafame, 
A.~C., Lanza, A.~F., Messina, S., 
\& Collier Cameron, A.\ 2011, \mnras, 416, 447 

%\bibitem[{{Spada}  {et~al.}(2011)}]{Spada2011}
%{Spada}, F.,  {Lanzafame}, A. C., \& {Lanza}, A.F. 2003, \mnras, 404, 641


\bibitem[{{Spruit}(2002)}]{Spruit2002}
{Spruit}, H.~C. 2002, \aap, 381, 923


\bibitem[{{Strugarek}  {et~al.}(2011a)}]{Strugarek2011a}
{Strugarek}, A.,  {Brun}, A. S., \& {Zahn}, J.P. 2011a, Astron. Nachricht., 332, 881


\bibitem[{{Strugarek}  {et~al.}(2011b)}]{Strugarek2011b}
{Strugarek}, A.,  {Brun}, A. S., \& {Zahn}, J.P. 2011b, \aap, 532, 34



\bibitem[{{Suijs} {et~al.}(2008){Suijs}, {Langer}, {Poelarends}, {Yoon},
  {Heger}, \& {Herwig}}]{Suijs2008}
{Suijs}, M.~P.~L., {Langer}, N., {Poelarends}, A.-J., {et~al.} 2008, \aap, 481, L87


\bibitem[{{Tayar} \& {Pinsonneault}(2013)}]{Tayar2013}
{Tayar}, J. \& {Pinsonneault}, M.H. 2013, \apj, 775, L1

\bibitem[Tayler(1973)]{Tayler73} Tayler, R.~J.\ 1973, \mnras, 
161, 365 

\bibitem[Tayler(1980)]{Tayler80} Tayler, R.~J.\ 1980, \mnras, 
191, 151 


\bibitem[{{Zahn}(1992)}]{Zahn92}
{Zahn}, J.-P. 1992, \aap, 265, 115


\bibitem[{{Zahn} {et al.}(2007)}]{ZahnBM2007}
{Zahn}, J.-P.,{Brun}, S.,{Mathis}, S. 2007, \aap, 474, 145


\end{thebibliography}
\end{document}